\documentclass[preprint,12pt]{elsarticle}
\usepackage{hyperref}
\usepackage{amssymb}
\usepackage{amsthm}
\usepackage{amstext}
\usepackage{amsmath}
\usepackage{amsfonts}
\usepackage{breqn}
\setkeys{breqn}{breakdepth={0}}
\usepackage{booktabs}
\usepackage{bbm}
\usepackage[german,english]{babel}
\usepackage{epstopdf}
\usepackage{multirow}
\usepackage{tikz}
\usepackage{verbatim}

\usepackage{listings}

\newcommand\red{\color{red}}
\newcommand\dif{{\rm d}}
\newcommand\preprint[1]{\gdef\@preprint{\hfill #1}}

\newcounter{bla}

\journal{Computer Physics Communications}

\usepackage{enumitem}
\usepackage{algorithm}
\usepackage{algorithmic}

\usepackage{tikz}
\usetikzlibrary{positioning}
\usetikzlibrary{arrows.meta}
\usetikzlibrary{shapes}
\usepackage{hyperref}
\usepackage{listings}
\usepackage{xcolor}
\usepackage{todonotes}
\definecolor{codegreen}{rgb}{0,0.6,0}
\definecolor{codegray}{rgb}{0.5,0.5,0.5}
\definecolor{codepurple}{rgb}{0.58,0,0.82}
\definecolor{backcolour}{rgb}{0.93,0.93,0.93}

\theoremstyle{plain}

\theoremstyle{definition}

\definecolor{ultramarine}{rgb}{0.0,0.28,0.68}

\usetikzlibrary{shapes,arrows}
\usetikzlibrary{decorations.pathmorphing}
\tikzstyle{startstop} = [rectangle,rounded corners=0.5cm, minimum width=2.5cm,minimum height=1cm,text centered, draw=black]
\tikzstyle{process} = [rectangle,rounded corners=0.2cm,minimum width=3cm,minimum height=1cm,text centered,text width=3cm,draw=black]
\tikzstyle{io} = [rectangle,minimum width=3cm,minimum height=1cm,text centered,text width =3cm,draw=black]
\tikzstyle{plaintext} = [rectangle,text centered,text width=3cm,minimum width=1cm,minimum height=0.5cm]
\tikzstyle{decision} = [trapezium, trapezium left angle = 70,trapezium right angle=110,minimum width=1cm,minimum height=0.5cm,text centered,text width=3cm,draw=black]
\tikzstyle{item} = [ellipse,minimum width=0.5cm,minimum height=1cm,text centered,text width=1cm,draw=black]
\tikzstyle{arrow} = [thick,->,>=stealth]
\tikzstyle{arrow2} = [thick,<->,>=stealth]

\begin{document} 
\preprint{USTC-ICTS/PCFT-23-15}

\begin{frontmatter}
\title{{\sc NeatIBP 1.0}, A package generating small-size
integration-by-parts relations for Feynman
integrals}
%\author[add1,add6]{Dominik Bendle}
%\ead{dominik.bendle@itwm.fraunhofer.de}
\author[add3,add4]{Zihao Wu}
\ead{wuzihao@mail.ustc.edu.cn}

\author[add1]{Janko Boehm}
\ead{boehm@mathematik.uni-kl.de}

%\author[add1,add6]{Murray Heymann}
%\ead{heymann@mathematik.uni-kl.de}

\author[add3]{Rourou Ma}
\ead{marr21@mail.ustc.edu.cn}
\author[add2]{Hefeng Xu}
\ead{rwinters@mail.ustc.edu.cn}

%\author[add6]{Mirko Rahn}
%\author[add1]{Lukas Ristau}
%\author[add1]{Marcel Wittmann}
%\ead{mwittman@rhrk.uni-kl.de}

\author[add3,add4,add5]{Yang Zhang}
\ead{yzhphy@ustc.edu.cn}

\address[add3]{Interdisciplinary Center for Theoretical Study, University of Science and Technology of China, Hefei, Anhui 230026, China}

\address[add4]{Peng Huanwu Center for Fundamental Theory, Hefei, Anhui 230026, China}

\address[add1]{Department of Mathematics, Rheinland-Pf\"alzische Technische Universit\"at Kaiserslautern-Landau (RPTU), 67663 Kaiserslautern, Germany}

\address[add2]{Department of Modern Physics, University of Science and Technology of China, Hefei, Anhui 230026, China}

\address[add5]{Max-Planck-Institut f\"ur Physik,
  Werner-Heisenberg-Institut, D-80805, M\"unchen, Germany}
%\address[add6]{Fraunhofer Institute for Industrial Mathematics
% (ITWM), Fraunhofer-Platz 1, 67663 Kaiserslautern, Germany}

\begin{abstract}
In this work, we present the package {\sc NeatIBP}, which automatically generates small-size integration-by-parts (IBP) identities for Feynman integrals. Based on the syzygy and module intersection techniques, the generated IBP identities' propagator degree is controlled and thus the size of the system of IBP identities is shorter than that generated by the standard Laporta algorithm. This package is powered by the computer algebra systems {\sc Mathematica} and {\sc Singular}, and the library {\sc SpaSM}. It is parallelized on the level of Feynman integral sectors. The generated small-size IBP identities can subsequently be used for either finite field reduction or analytic reduction. We demonstrate the capabilities of this package on several multi-loop IBP examples.
\end{abstract}


\end{frontmatter}

\pagebreak[4]
\noindent {\bf PROGRAM SUMMARY}\smallskip

\begin{small}
\noindent
{\em Manuscript Title:} {{\sc NeatIBP 1.0}, A package for generating small-size
integration-by-parts relations for Feynman integrals}\\
{\em Authors:} {Zihao Wu, Janko Boehm, Rourou Ma, Hefeng Xu, Yang Zhang}\\
{\em Program Title:}   {NeatIBP}                                      \\
{\em CPC Library link to program files:} (to be added by Technical Editor) \\
{\em Developer's repository link:} \url{https://github.com/yzhphy/NeatIBP}\\
{\em Code Ocean capsule:} (to be added by Technical Editor)\\
{\em Licensing provisions(please choose one):} {GPLv3}  \\
{\em Programming language:}       {\sc Mathematica}, {\sc Singular}, {\sc C}                  \\
{\em Computer(s) for which the program has been designed:} computers with multiple CPU cores\\
{\em Operating system(s) for which the program has been designed:} {Linux}   \\
{\em Supplementary material:} {none}                                \\
  % Fill in if necessary, otherwise leave out.
{\em Nature of problem:} {The difficulty of the IBP reduction of Feynman integrals stems from the large size of the IBP system.}
\\
{\em Solution method:} {We apply the module intersection method to generate IBP relations by algebraic geometry methods, and compare the smaller IBP system we generate with that from Laporta's algorithm. {\sc NeatIBP} generates the module intersection in {\sc Singular} \cite{1} and selects relevant and independent IBP relations via an algorithm relying on the library {\sc SpaSM} \cite{2}. The workflow is parallelized by a task manager written in {\sc Mathematica} \cite{3}.  }
  %Describe the method solution here.
%{\em Additional comments including restrictions and unusual features (approx. 50-250 words):}{\color{red} comments}
  %Provide any additional comments here.
   \\

\end{small}

\section{Introduction}
\label{sec:intro}

Feynman integrals are the core objects in perturbative quantum field theory. They are crucial in the theoretical computations, especially which is related to the observations of high energy experiments, such as, in Large Hadron Collider (LHC) \cite{Evans:2008zzb,LHCb:2008vvz,CMS:2008xjf,ATLAS:2008xda}. As the increasing of precision in high-energy experiments, the demand for multi-loop Feynman integral computations going to rise continuously. Typically, for a given scattering process, the computation of scattering amplitude involves numerous Feynman integrals' contribution. One thus performs reductions \cite{Passarino:1978jh,Tkachov:1981wb,Chetyrkin:1981qh,Ossola:2006us,Ossola:2007ax} on the integrals towards smaller sets of integrals which then have to be computed. 

Integration-by-Parts (IBP) reduction \cite{Tkachov:1981wb,Chetyrkin:1981qh} is a key Feynman integral reduction technique. %It is a critical step to reduce large number of scalar Feynman integrals into small amount of master integrals. 
Laporta's algorithm \cite{Laporta:2000dsw} is the most prevalent IBP reduction algorithm, relying on reducing a linear system of IBP relations. A multitude of other approaches exist for executing IBP reduction, including the use of operators and non-commutative algebra \cite{Smirnov:2006wh,Smirnov:2005ky,Bitoun:2017nre,Lee:2008tj,Lee:2014tja,Baikov:1996iu,Grozin:2011mt,Barakat:2022qlc}, intersection numbers \cite{Mastrolia:2018uzb,Frellesvig:2019uqt,Frellesvig:2019kgj,Frellesvig:2020qot}, auxiliary mass flow methods \cite{Liu:2017jxz,Liu:2018dmc,Guan:2019bcx,Zhang:2018mlo,Wang:2019mnn,Liu:2022mfb,Liu:2022chg,Liu:2021wks,Liu:2020kpc}, finite field methods \cite{vonManteuffel:2014ixa,Peraro:2016wsq,Klappert:2019emp,Klappert:2020aqs,Peraro:2019svx}, methods of direct solutions \cite{Kosower:2018obg}, and more. Building on these techniques, there are several packages for IBP reductions available, including AIR \cite{Anastasiou:2004vj}, FiniteFlow\cite{Peraro:2019svx}, FIRE \cite{Smirnov:2008iw,Smirnov:2013dia,Smirnov:2014hma,Smirnov:2019qkx}, {\sc Kira} \cite{Maierhofer:2017gsa,Maierhofer:2018gpa,Maierhofer:2019goc}, {\sc LiteRed} \cite{Lee:2013mka}, and {\sc Reduze} \cite{Studerus:2009ye,vonManteuffel:2012np}.  % with the Feynman involved integrals ordered by some chosen criterion.  commercial algebraic software.

One of the key factors determining the computational complexity of IBP reduction is the number of initial IBP relations.
%This challenge occurs especially in some cutting-edge Feynman %diagrams. For these diagrams, the IBP reduction process serves as %a bottle neck. Thus, we need to ask such a question: \textit{Is %there any way to make the IBP reduction process less complicated?}
%Yes, there is. %do we need to introduce other way?
%As stated above, the hardness of the IBP reduction relies on the size of the coefficient matrix, or equivalently, the numbers of the IBP relations and in which the Feynman integrals involved. 
%For a given Feynman integrals reduction problem, there are different kinds of methods to generate different linear system of IBP relations, while the sizes of the systems can be very differently. This stimulates us to find a good method that generates relatively small-sized IBP system.
%Thus, one answer for the question raised above is to find a good method that generates relatively small-sized IBP system for a given reduction problem. 
The syzygy method for IBP reduction \cite{Gluza:2010ws} is a powerful method to decrease this number of IBP relations. So-called syzygy equations impose constraints on the Feynman integral indices, and thus significantly decrease the number of IBP relations to consider in subsequent computations. Several novel methods have been developed for generating or reformulating syzygy equations, including linear algebra techniques \cite{Schabinger:2011dz,10.1145/1993886.1993902}, independent syzygy generator selection \cite{Chen:2015lyz} and module intersection \cite{Bohm:2017qme,Bosma:2018mtf,Boehm:2020zig}. %Although there are a lot of development in syzygy IBP methods, {\it an automatic syzygy-method-based IBP program seems rare on the market.}
Moreover, syzygy method can also be used for generating IBP relations of specific type Feynman integrals, such as Feynman integrals with only overall UV divergence \cite{Henn:2022vqp}. 
Although there have been numerous advancements in syzygy methods for IBP reduction, it appears that {\it an automated program employing syzygy method-based IBP techniques is noticeably lacking in the market.
}
%By preventing the integrals' propagator power from being increased, syzygy method makes the IBP system much smaller than traditional methods. This results in a much higher efficiency for the IBP reduction process. 

In this paper, we introduce our package {\sc NeatIBP} (ver. 1.0). The tool automatically generate small-size IBP systems, relying on syzygy techniques and, in particular, the module intersection method \cite{Bohm:2017qme,Bosma:2018mtf}. 

In our package, we employ the Baikov representation of Feynman integrals \cite{Baikov:1996cd,Baikov:1996rk,Baikov:2005nv}. The index restriction condition of Feynman integrals in IBP relations, translates into a module intersection computation over the polynomial ring in the variables of the Baikov representation \cite{Bohm:2017qme,Bosma:2018mtf}. The module intersection in the current version of {\sc NeatIBP} is determined using the computer algebra system {\sc Singular} \cite{DGPS}. Then formal IBP relations\footnote{In this paper, \textit{formal} (IBP) relations refer to IBP relations of integrals with symbolic indices.}, with restriction of indices, are generated. Based on the formal relations, we introduce sophisticated seeding\footnote{ \textit{Seeding} is the procedure of setting symbolic indices in formal IBP relations as specific numbers.} and selection algorithms to generate a small size IBP system. During the seeding steps, careful control is exercised over the numerator degree of the generated IBP relations.  Using row reduction methods over finite fields\footnote{One should keep in mind that row reduction over finite fields always has risks. An ill-chosen numeric point or prime number could lead to a wrong answer. We will discuss this in Section \ref{subsec:Generating IBP relations in one sector}.}, a small set of sufficient IBP relations is selected. Along our way, the master integrals are determined.
The selection step is powered by the finite field linear system reduction package {\sc SpaSM} \cite{spasm}. 

The output of NeatIBP is a list of IBP relations, which are sufficient to reduce input integrals to master integrals. This list, which is typically much shorter than the one needed in Laporta's approach, serves two purposes:
\begin{enumerate}
    \item Integral reduction over finite fields, with a subsequent reconstruction multi-loop amplitudes.
    \item Analytic integral reduction, if required.
\end{enumerate}
The user can use their own programs to finish the above two approach in the current stage. In the future, we will develop corresponding interfaces, feeding {\sc NeatIBP} results into some mainstream IBP reduction or Gaussian elimination software automatically.

{\sc NeatIBP} 1.0 is employing parallelization. %Usually, a reduction task can be divided into multiple smaller sub tasks by the \textit{sectors} of the involved integrals. The sector of a Feynman integral describes which propagators appear in the denominator. The sectors of the integrals may have some sub-super relations. 
In the initial step, the zero sectors and symmetric sectors are determined using standard methods \cite{Lee:2013mka}. The symmetry maps between sectors are found in a parallelized manner. Subsequently, the module intersection, IBP relation seeding, and selection computations are parallelized across distinct Feynman integral sectors. The status of the computation is visible in our monitor program. Important intermediate results are stored in real-time. Due to this approach, the execution of the program can be interrupted and resumed. 

%In {\sc NeatIBP}, IBP relations lead by integrals in a sector can be generated together. The generation of IBP relations on different sectors can be complied simultaneously, if they do not have any super-sub relations between each other. 

%By a well-designed mission distribution system between the sectors, {\sc NeatIBP} arranges the generation tasks with the efficiency as much as possible.  {\color{red} (may need to rewrite this sentence)}

This paper is organized as follows. In Section \ref{sec:methods}, we provide an overview of the mathematical physics methods used in our program, and briefly outline their implementations in {\sc NeatIBP}. In Section \ref{sec:algorithm}, we present the underlying algorithm of {\sc NeatIBP}, as well as its parallelization scheme. In Section \ref{sec:manual} provides a short guide on how to use {\sc NeatIBP}, with more details available online. In Section \ref{sec:examples}, we showcase examples for the usage {\sc NeatIBP} and give timings. Section \ref{sec:summary} summarizes the paper.  

%We would like to state about the author ranking of this paper. 

%Conventionally, the authors names of this paper should have been arranged in a lexicographic ordering. 

We remark that due to graduation requirements of the University of Science and Technology of China necessitating first-author papers, and Wu's significant contribution to this work, the author names ordering of this paper is chosen in the stated way.

%===================================================================

\section{Review of methods and their implementations}\label{sec:methods}
\subsection{IBP reduction}
IBP reduction refers to reducing given scalar integrals in a Feynman integral family to a finite number of linearly independent integrals. A Feynman integral family, for a given Feynman diagram, is a set of Feynman integrals of the form
\begin{equation}\label{eq:FI}
    G_{\alpha_1,\cdots,\alpha_n}=\int\prod_{j=1}^{L}\frac{\dif^D l_j}{i \pi^{D/2}}\frac{1}{D_1^{\alpha_1} \cdots D_n^{\alpha_n}},
\end{equation}
where the $D_i$ are the propagators associated to the Feynman diagram, $\alpha_i$'s are the power indices, usually taking integer values. The IBP relations originate from the vanishing boundary term in dimensional regularization, such that 
\begin{equation}\label{eq:momentum IBP}
    0=\int\prod_{j=1}^{L}\frac{\dif^D l_j}{i \pi^{D/2}}\frac{\partial}{\partial l_k^\mu}\frac{v^\mu}{D_1^{\alpha_1} \cdots D_n^{\alpha_n}},
\end{equation}
where $v^\mu$ is a vector formed by loop or external momenta. The right hand side of \eqref{eq:momentum IBP} can be expanded as a linear combination of multiple Feynman integrals in this family. Thus, \eqref{eq:momentum IBP} generates infinite number of linear constrains on the Feynman integrals in the family. Constrained by the relations, only finite \cite{Smirnov:2010hn} number of Feynman integrals in the family are independent. They can be chosen to form a linear basis. Any integral $I$ in the family can be represented as 
\begin{equation}\label{eq:IBP reduction}
    I = \sum_{i}c_i I_i,
\end{equation}
where $I_i$'s are called \textit{master integrals}, $c_i$'s are the reduction coefficients.
In order to reduce some target integrals toward a chosen master integral basis, we usually follow the following steps. Firstly, we generate enough IBP relations from \eqref{eq:momentum IBP}. %The relations are some linear equations. 
We then need to solve the linear equation system, with the target integrals treated as the unknown and master integrals as the known. Finally we will get the reduction result \eqref{eq:IBP reduction}.

Usually, \eqref{eq:momentum IBP} generates IBP relations with the indices $\alpha_i$'s (for denominators) increased, for a generally chosen $v^\mu$. This phenomenon will significantly increase the size of the IBP system, since the integrals with increased indices are often irrelevant to the reduction target. Fortunately, new methods can avoid this the appearance of such unneeded integrals, by solving syzygy equations \cite{Gluza:2010ws,Chen:2015lyz,Larsen:2015ped,Zhang:2016kfo,Schabinger:2011dz,Bohm:2017qme,Bosma:2018mtf,Boehm:2020zig}. In {\sc NeatIBP}, we use IBP relations from Baikov representation, with the help of syzygy equations and the module intersection method\cite{Bohm:2017qme,Boehm:2020zig}. We will introduce them in the subsequent sections.

%-----------------------------------------------------------

\subsection{IBP relations from the Baikov representation}\label{subsec:IBP relations from the Baikov representation}

The Baikov representation of Feynman integrals is 
\begin{equation}\label{eq:Baikov rep}
    G_{\alpha_1,\cdots,\alpha_n}=C\int\dif z_1 \cdots \dif z_n P^\gamma \frac{1}{z_1^{\alpha_1} \cdots z_n^{\alpha_n}},
\end{equation}
where $C$ is a constant irrelevant to the IBP relations, $\gamma=\frac{D-L-E-1}{2}$, with $L$ and $E$ the number of independent loop and external momenta respectively, and 
\begin{equation}
    P=P(z_1,\cdots,z_n)=\det G(l_1,\cdots,l_L,p_1,\cdots p_E),
\end{equation}
standing for the Gram determinant of loop and external momenta. 

In the Baikov representation, IBP relations are generated from the total derivative expression as
\begin{equation}\label{eq:Baikov total direvative}
    0=C\int\dif z_1 \cdots \dif z_n \sum_{i=1}^n \frac{\partial}{\partial z_i} \Big(a_i(z) P^\gamma \frac{1}{z_1^{\alpha_1} \cdots z_n^{\alpha_n}}\Big),
\end{equation}
where $a_i(z)$'s are some chosen polynomials of $z_j$'s. If one can find such $a_i$'s that there exist corresponding polynomials $b(z)$ and $b_i(z)$'s satisfying
\begin{equation}\label{eq:Baikov syzygy}
    b(z) P + \sum_{i=1}^n \Big(a_i(z)\frac{\partial P}{\partial z_i}\Big)=0, 
\end{equation}
and
\begin{equation}\label{eq:Baikov index constrian}
    a_i(z)=b_i(z) z_i, \quad \text{for } i\in \{j|\alpha_j>0\},
\end{equation}
\eqref{eq:Baikov total direvative} yields,
\begin{equation}\label{eq:Baikov IBP}
    0=C\int\dif z_1 \cdots \dif z_n \Big(\sum_{i=1}^n\big( \frac{\partial a_i}{\partial z_i} -\alpha_i \frac{ a_i}{ z_i}\big)-\gamma b\Big) P^\gamma \frac{1}{z_1^{\alpha_1} \cdots z_n^{\alpha_n}},
\end{equation}
which generates IBP relations via the Baikov representation, keeping the \textit{non-negative} indices from being increased. We call these IBP relations \textit{formal IBP relations} if the indices $\alpha_i$ are  symbolic.

The equations \eqref{eq:Baikov syzygy} are the syzygy equations of the set of generators \\$\langle P, \frac{\partial P}{\partial z_1},\cdots,\frac{\partial P}{\partial z_n}\rangle$. Considering Baikov kernel $P$ is defined as the determinate of certain matrix, which let us call as $P_{ij}$, we have such a Laplace expansion \cite{Bohm:2017qme} as
\begin{equation}\label{eq:Laplace expansion}
    \sum_j P_{ij}\frac{\partial P}{\partial P_{kj}}-\delta_{ik} P=0.
\end{equation}
The matrix entries, $P_{ij}$, are scalar products of the momenta, thus being linear combinations of the propagators $z_i$'s and kinematic variables. Consequently, from the Laplace expansion \eqref{eq:Laplace expansion}, the solutions of the syzygy equation \eqref{eq:Baikov syzygy} can be directly read, using the chain rule of derivative. The completeness of those syzygy generators are proved in \cite{Bohm:2017qme}.

Given a Baikov kernel $P$, the solutions of \eqref{eq:Baikov syzygy}, which are vectors of polynomials, $(b,a_1,\cdots,a_n)$, are not unique. All the solutions form an algebraic structure called \textit{module}. Call the solution module of \eqref{eq:Baikov syzygy} $M_1$. Similarly, the solutions of \eqref{eq:Baikov index constrian} also form a module, $M_2$. In order to find the common solutions of \eqref{eq:Baikov syzygy} and \eqref{eq:Baikov index constrian}, we perform \textit{module intersection} \cite{Boehm:2020zig}
\begin{equation}\label{eq:module intersection}
    M=M_1 \cap M_2\,.
\end{equation}
In the current version of \textsc{NeatIBP}, the module intersection problem is solved using syzygy method. Let us denote the generators of $M_1$ and $M_2$ as $f_i$ and $g_i$, respectively. We solve the syzygy equation of $\langle f_i \rangle \cup \langle g_i \rangle$ to get
\begin{equation}
\sum_i s_i f_i+\sum_j r_j g_j=0.
\end{equation}
Then, $\sum_i s_i f_i$ is both in $M_1$ and $M_2$, serving as a generator of $M_1\cap M_2$.

From the module intersection \eqref{eq:module intersection}, we  get the generators of module $M$. They can be used in \eqref{eq:Baikov IBP} to generate formal IBP relations in the following form,
\begin{equation}\label{eq:FIBP}
    0=\sum_{\{\vec{\beta}\}} c_{\{\vec{\beta}\}}(\alpha_1,\cdots,\alpha_n)I_{\alpha_1+\beta_1,\cdots,\alpha_n+\beta_n}\,.
\end{equation}
With the power indices $\alpha_i$ in the formal IBP relations as integers, specific IBP relations are obtained. This process is called \textit{seeding}, and the specific integer value for $\alpha_i$'s is called a \textit{seed}.

Seeding is a key step for many IBP reduction algorithms. For syzygy or module intersection-based IBP reduction, the seeding process would be more sophisticated since the syzygy/module intersection generators may provide IBP relations with different numerator degrees. The details of seeding in {\sc NeatIBP} v1.0 is described in the next section.

%---------------------------------------------------------
\subsection{The sectors of a Feynman integral family}
 The integrals in an integral family can be categorized by the propagators that appear in the denominator of the integrand. Such a category is called a \textit{sector} in the integral family. An integral $G_{\alpha_1,\cdots,\alpha_n}$ is in a sector  $(s_1,\cdots,s_n)$, if,
\begin{equation}\label{eq:sector def}
    s_i=
    \left\{
        \begin{array}{lr}
            1,\quad\alpha_i>0\\
            0,\quad\alpha_i\leq0
        \end{array}
    \right.
    ,
\end{equation}
%where ``1'' stands for that the corresponding propagator appears in the denominator and ``0'' stands for the otherwise. 
%There are some super-sub relations between the sectors. 
Changing one or more ``1'' to ``0'' in the sector expression $(s_1,\cdots,s_n)$ of a certain sector $A$ , %which stands for removing one propagator in the denominator of the corresponding integrals, 
we get another sector $B$ which is called a \textit{sub sector} of sector $A$. Accordingly, sector $A$ is called a \textit{super sector} of sector $B$. Sometimes, one can label a sector by %the positions of the propagators that appears in the denominator, which is 
the set $\{i|s_i=1\}$. For example, sector $(0,1,0,0,1,0,1,0,0)$ can be labeled as $\{2,5,7\}$.

%\subsubsection{The zero sectors}
Among the $2^n$ sectors in an integral family with $n$ propagators, there are many of them called \textit{zero sectors}, in which all integrals are zero.  {\sc NeatIBP} applies Lee's zero sector criteria to determine zero sectors \cite{Lee:2013mka}. %We would like introducing Lee's zero sector criteria briefly here. 
This method is based on Lee-Pomeransky representation \cite{Lee:2013mka,Lee:2013hzt,Lee:2014tja}. %By considering the scale transformation in Lee-Pomeransky representation, we can determine whether a sector is a zero sector. 
It is shown in \cite{Lee:2013mka} that for a given sector $S$, if there exist constants $t_i$'s  such that
\begin{equation} 
\label{eq: criterion of zero sector}
    \sum_i t_i x_i \frac{\partial G}{\partial x_i}=G,
\end{equation}
where $G$ denotes the Lee-Pomeransky polynomial on sector $S$, then $S$ is a zero sector. %In {\sc NeatIBP}, we are using this method to determine zero sectors.
%A zero sector is a scaleless sector indeed, in Lee-Pomeransky representation \cite{Lee:2013mka,Lee:2013hzt,Lee:2014tja}, a scaleless integral should be proportional to itself before the integral variables substitution $x_i \to (1+t_i \omega)x_i$, say,

%\begin{equation}
%    G_{\alpha_1, \cdots,\alpha_n}=(1+\omega)G_{\alpha_1,\cdots,\alpha_n}
%\end{equation}
%where $\omega$ is infinitesimal parameter, and $t_i$ are some finite coefficients. Or equivalently, to Lee's criterion of zero sector: there exist a solution $t_i$ for Eq. \eqref{eq: criterion of zero sector}, if $G_{\alpha_1,\cdots,\alpha_n}$ is in zero sector.

%where $G$ denotes the Lee-Pomeransky polynomial $G=U+F$ in this sector and $t_i$ are constants irrelevant to $x_i$.
%The zero sector detection part of our package is written in {\sc Mathematica}. Zero sectors will not appear in the symmetry search or the IBP generating steps. 
%If a input integral $G_{\alpha_1,\cdots,\alpha_n}$ is in one zero sector, the output will contain a relation $G_{\alpha_1,\cdots,\alpha_n}=0$ to mark it.

%----------------------------------------------------------------------
\subsection{The symmetries of Feynman integrals}\label{subsec: symmetry review}
%{\red We need to FURTHER shrink this sub section.}

Symmetries between Feynman integrals are important for the integral reduction.

%we must consider, since this kind of symmetries, which originate from the topology of Feynman diagrams, are sometimes independent of IBP relation. %, which remain unchanged after permuting some of its propagators. The symmetry relations are sometimes independent of IBP relations, so we must consider them additionally. 

It is well-known that the symmetries of Feynman integrals can be found from the polynomial symmetry of their parameterized representations. The polynomial symmetries can be found by Pak's algorithm \cite{Pak:2011xt}. In this way, the symmetry between Feynman integrals via swapping propagators is determined.

%We can get a symmetry relation by some integral variables’ substitution $good enough to transform a Feynman integral to another Feynman integral $in form \eqref{eq:FI}, $which means these two Feynman integrals share a same set of propagators. Therefore, the integral variables substitution
Furthermore, the propagator permutation can be reformulated to a momentum map, which is found by the following ansatz,
\begin{align}
\label{eq:momentum symmetry internal trans.}
    l_i\mapsto A_{ij} l_j + B_{ik}p_k, &\quad \text{where} \, \det(A_{ij})=\pm 1,
    &p_i\mapsto C_{ij} p_j.
\end{align}
corresponding to the permutation,
\begin{equation}\label{eq:momentum symmetry trans prop. condition}
    D_i\to D_{\sigma(i)}, \quad\text{for }i\in\{j|\alpha_j>0\},
\end{equation}
where $\sigma$ stands for a certain permutation. We call a symmetry relation a \textit{symmetry inside a sector}, if the permutation rule $\sigma$ map integrals in one sector to integrals in the same sector. We call it a \textit{symmetry between sectors} otherwise. The coefficients in the ansatz \eqref{eq:momentum symmetry internal trans.} are determined by the propagator matching and the invariance of the Gram matrix of external legs. 

We remark that, a momentum map is stronger than the polynomial symmetry. For a polynomial symmetry, in some cases, the corresponding momentum map does not exist. There may be other cases that it does exist, but some efforts are needed to derive it. For instance, {\sc NeatIBP} will search symmetry relations in lower sectors by redefining external momenta.

{\subsection{Propagator cuts}}
For complicated integrals family, frequently one would like to carry out the IBP reduction with propagator cuts. For integrals with Dirac delta functions in the integrand, IBP relations with cuts are also necessary. In {\sc NeatIBP v1.0}, computation with propagator cut is also included. For the conceptual introduction of the propagator cuts and the related generalized unitarity methods, we refer to the ref.~ \cite{Bern:1994cg}\cite{Berger:2008sj}\cite{Britto:2004nc}\cite{Larsen:2015ped}. 

In this version of our package, we implement the propagator cut convention in the Baikov representation as \cite{Larsen:2015ped}. Suppose that $\mathcal C$ is the set of propagator indices on the cut, then the Baikov integral kernel becomes,
\begin{equation}
    P \mapsto P|_{z_j\to 0}, \quad j\in \mathcal C\,.
\end{equation}
and the IBP relation can be found by the following module intersection,
\begin{equation}
    \big(M_1|_{z_j\to 0}\big) \cap \big(M_2|_{z_j\to 0}\big),  \quad j\in \mathcal C\,.
\end{equation}
The resulting IBP relations have no increase of the uncut indices of propagators. Usually, with a cut, the module intersection computation and the IBP seeding would be much easier. For complicated Feynman integral family, it makes sense to consider all the spanning cuts of the IBP reduction and then construct the full IBP reduction. 

In this version of {\sc NeatIBP}, we only deal with the Feynman integrals with the propagator index lower than $2$ {\it on the cut}. The integrand for double propagators on the cut in the Baikov representation, which contains the derivative of the kernel $P$ is more complicated. We leave this case for the future versions.

We also remark that the choice of propagator cuts usually breaks the symmetry between sectors. So, in this version of our package, if the propagator cut is imposed then the symmetry searching should be turned off.

%=======================================================================

\section{Algorithms} \label{sec:algorithm}
\begin{comment}

\if 0
\subsection{Overview}
\begin{enumerate}[label=(\arabic*)]
    \item Validate the input.
    \item List all sectors of the input target integrals, as well as their sub sectors. These sectors form the relevant sectors of this problem.
    \item Find out zero sectors. Save all target integrals in the zero sectors as $0$ in the output IBP relation file.
    \item If the symmetry option is on, find symmetries between sectors, and build up a rule of sector mapping. This will divide all nonzero sectors into two class: mapped sectors and unique sectors.
    \item If the symmetry option is on, convert all target integrals in the mapped sectors using symmetry map relations, and save the corresponding identity in the exporting IBP relation files.
    \item Build up the sector relation tree.
    \item Initialize the mission status of each sector as ``waiting super sectors''.
\end{enumerate}
In this subsection, we are going to {\red ...}. Detailed algorithms of each step will be introduced later.
The goal of our package is to reduce the given target integrals in a certain integral family. The integral family can be divided into different sectors. Firstly, we need to find out symmetries between sectors. 
\begin{figure}[h]
\centering
\includegraphics[width=1.035\textwidth]{sector_mapping_with_sub_sectors.eps}
\caption{sector mapping with sub sectors}
\label{fig:sector mapping with sub sectors}
\end{figure}
\fi 
\end{comment}
In this section, we will introduce the main ideas of the algorithm of {\sc NeatIBP}. 

\subsection{Generating IBP relations referring to sectors}\label{subsec:Generating IBP relations referring to sectors}
In {\sc NeatIBP}, the IBP reduction relations are generated by seeding process referring to the sector structure of an integral family.
%{\red connection words}
%The idea of sectors can be generalized to the seeding of formal IBP relations. 
%A specific list of integer values for $\alpha_i$'s is called a {\it seed}. %We can also define the sector of a seed by \eqref{eq:sector def}. 
%Seeding formal IBP relations from \eqref{eq:Baikov IBP} using a seed $s$ from a certain sector will result in  IBP relations that only involve integrals in the current sector and the sub sectors. %Thus, the specific IBP relations resulting from the seeding process are also with a structure labelled by sectors. 
%In the module intersection method of generating IBPs, 
We apply the similar idea ``tail mask" as that in the Laporta algorithm. For each sector, we just accumulate IBP relations which can reduce the target integrals to the master integrals of this sector, and the (unreduced) integrals in the sub sectors. The integrals from the sub sectors are also called \textit{tail integrals}\footnote{In the following discussions, we call integrals from the current sectors as \textit{head integrals}.}.
%The main idea of generating IBP relations referring sector structures is that, while seeding on a given sector, we treat the integrals on current sectors as unknowns of the linear equations formed by IBP relations, but the integrals from the sub sectors as knowns. 
%The integrals are also called \textit{tail integrals}.\footnote{In the next paragraphs, we call the integrals from the current sectors as \textit{head integrals}.} %will be treated as target integrals of the corresponding sub sectors they are in. %And in the seeding of the sub sectors, fine IBP relations will be generated to solve these integrals treated as the target integrals in the subsectors, as unknowns there. 
\begin{comment}

The IBP relations generated from a sector should satisfy the follow requirements:
\begin{itemize}
    \item Being enough: they reduce all target integrals in the corresponding sectors as linear combinations of master integrals in the current sector and (unreduced) tail integrals. 
    \item Being linearly independent: No linear combination of this IBP system may give an IBP relation that \textbf{only} involves tail integrals.
    
\end{itemize}
    
\end{comment}
%Equivalent speaking, the IBP system must be enough\footnote{This means being enough to reduce all target integrals as linear combinations of master integrals on this sector.} and linearly independent after setting all the tail integrals to 0. 

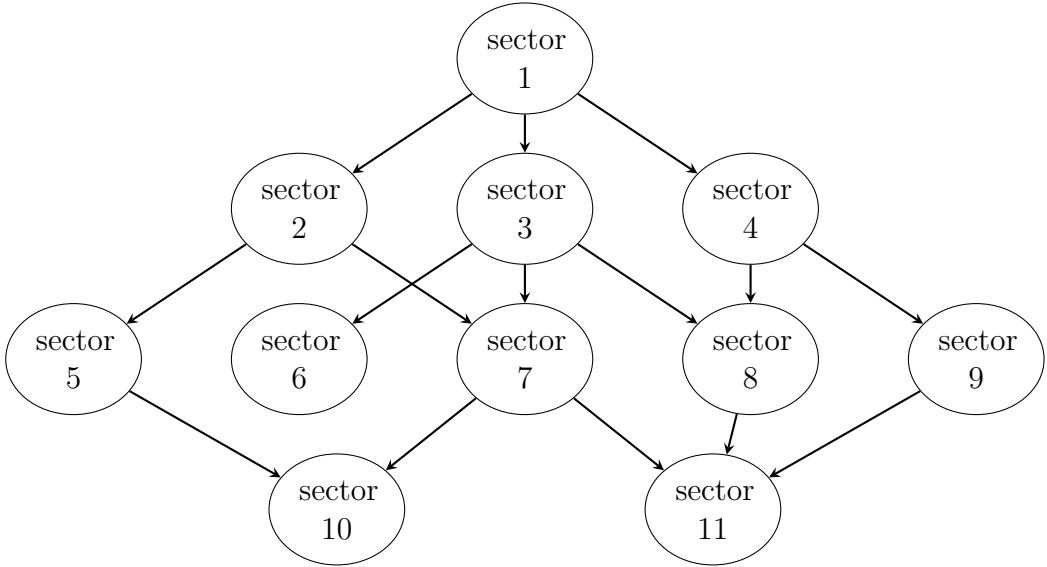
\begin{figure}[htb]
\centering
\begin{comment}
\begin{tikzpicture}[node distance=2cm,scale=0.7]
\node (1) [item] {sector 1};
\node (2) [item,below of=1,xshift=-3cm ] {sector 2};
\node (3) [item,below of=1,xshift=0cm ] {sector 3};
\node (4) [item,below of=1,xshift=3cm ] {sector 4};
\node (5) [item,below of=2,xshift=-3cm ] {sector 5};
\node (6) [item,below of=2,xshift=0cm ] {sector 6};
\node (7) [item,below of=2,xshift=3cm ] {sector 7};
\node (8) [item,below of=4,xshift=0cm ] {sector 8};
\node (9) [item,below of=4,xshift=3cm ] {sector 9};
\node (10) [item,below of=7,xshift=-2.5cm ] {sector 10};
\node (11) [item,below of=7,xshift=2.5cm ] {sector 11};
\draw [arrow] (1) -- (2);
\draw [arrow] (1) -- (3);
\draw [arrow] (1) -- (4);
\draw [arrow] (2) -- (5);
\draw [arrow] (3) -- (7);
\draw [arrow] (2) -- (7);
\draw [arrow] (3) -- (8);
\draw [arrow] (3) -- (6);
\draw [arrow] (4) -- (8);
\draw [arrow] (4) -- (9);
\draw [arrow] (5) -- (10);
\draw [arrow] (7) -- (10);
\draw [arrow] (9) -- (11);
\draw [arrow] (8) -- (11);
\draw [arrow] (7) -- (11);
\end{tikzpicture}
\end{comment}

\begin{tikzpicture}[node distance=3cm,scale=0.7]
\node (118) [item] {sector 118};
\node (116) [item,below of=118,xshift=-4.5cm ] {sector 116};
\node (114) [item,below of=118,xshift=-1.5cm ] {sector 114};
\node (102) [item,below of=118,xshift=1.5cm ] {sector 102};
\node (86) [item,below of=118,xshift=4.5cm ] {sector 86};
\node (100) [item,below of=114,xshift=0cm ] {sector 100};
\node (82) [item,below of=102,xshift=0cm ] {sector 82};

\draw [arrow] (118) -- (116);
\draw [arrow] (118) -- (114);
\draw [arrow] (118) -- (102);
\draw [arrow] (118) -- (86);
\draw [arrow] (116) -- (100);
\draw [arrow] (102) -- (100);
\draw [arrow] (86) -- (82);
\draw [arrow] (114) -- (82);

\end{tikzpicture}
\caption{An example of a web structure of the sectors in a slashed box family, as a sub family of the double box.}
\label{fig: sector web structure}
\end{figure}

%Adapting the above ideas, 
Generating IBP relations in a family is divided into sub tasks which correspond to the sectors. The family can be visualized as a web structure with each sector as a node, (see Fig. \ref{fig: sector web structure}). The input of each node is a set of the target integrals of the current sector, consisting user input and tail integrals from its super sectors. The output of each node includes master integrals and IBP relations generated from the current sector. Tail integrals in these IBP relations would become the input of the corresponding sub nodes. 

\begin{figure}
    \centering
    \begin{tikzpicture}[scale=0.8]
        \node(manager) [startstop] at (0,3) {task manager};
        \node(table) [io] at (0,0) {registration table of the mission statuses};

        \draw[->] (manager) edge[bend left=30] node[below,xshift=1.2cm,font=\fontsize{8}{16}\selectfont] {modify statuses} (table);
  \draw[->] (table) edge[bend left=30] node[above,xshift=-0.9cm,font=\fontsize{8}{16}\selectfont] {read statuses} (manager);

  \node(worker 1) [startstop] at (-4,-3) {worker 1};
  \node(worker 2) [startstop] at (0,-3) {worker 2};
  \node(worker i) [plaintext] at (4,-3) {$\cdots$};
\node(data) [io] at (0,-6) {data storage};

 \draw[->] (worker 1) edge[bend right=10] node[below,xshift=0.83cm,yshift=0.7cm,font=\fontsize{8}{4}\selectfont,text width=1cm] {report finish} (table);
  \draw[->] (table) edge[bend right=10] node[above,xshift=-0.85cm,yshift=-0.7cm,font=\fontsize{8}{4}\selectfont,text width=1.5cm] {send start signal} (worker 1);
   \draw[->] (worker 2) edge[bend right=10] node[below,xshift=0.55cm,yshift=0.7cm,font=\fontsize{8}{4}\selectfont,text width=1cm] {report finish} (table);
  \draw[->] (table) edge[bend right=10] node[above,xshift=-0.6cm,yshift=-0.7cm,font=\fontsize{8}{4}\selectfont,text width=1.5cm] {send start signal} (worker 2);
   \draw[->] (worker i) edge[bend right=10]  (table);
  \draw[->] (table) edge[bend right=10]  (worker i);

  \draw[->] (worker 1) edge[bend left=10] node[below,xshift=0.45cm,yshift=0.33cm,font=\fontsize{8}{4}\selectfont,text width=1cm] {outputs} (data);
  \draw[->] (data) edge[bend left=10] node[above,xshift=-0.7cm,yshift=0cm,font=\fontsize{8}{4}\selectfont,text width=1.5cm] {inputs} (worker 1);
   \draw[->] (worker 2) edge[bend left=10] node[below,xshift=0.52cm,yshift=0.1cm,font=\fontsize{8}{4}\selectfont,text width=1cm] {outputs} (data);
  \draw[->] (data) edge[bend left=10] node[above,xshift=-0.1cm,yshift=0.32cm,font=\fontsize{8}{4}\selectfont,text width=1.5cm] {inputs} (worker 2);
   \draw[->] (worker i) edge[bend left=10]  (data);
  \draw[->] (data) edge[bend left=10]  (worker i);

    \end{tikzpicture}
    \caption{Distributed computation based on registration table.}
    \label{fig:Distributed computation based on registration table}
\end{figure}
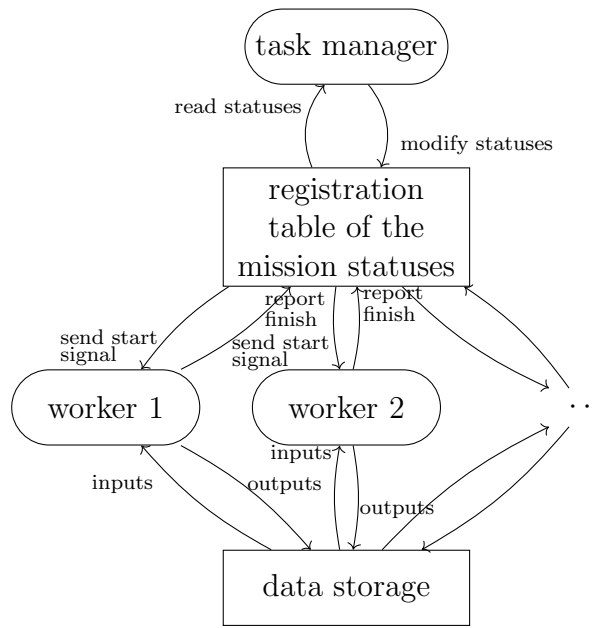

The web structure enables us to distribute the computation in parallel. In our parallelization scheme, %considering the input and output feature in this web, 
a task immediately starts once all its super tasks are finished.
%A parallelization management is required in the implementation of this. 
In {\sc NeatIBP} v1.0, the parallel tasks are arranged by a mission status registration table and a task manager. Logically, such a parallelization scheme is built in a way shown in Fig. \ref{fig:Distributed computation based on registration table}. In the registration table, the mission statuses have four cases
%\footnote{The wording of the statuses here are not identical to those in the source codes of {\sc NeatIBP}, but they represents the same meaning.}
. They are \textit{waiting}, \textit{ready to compute}, \textit{computing} and \textit{finished}.
%\begin{itemize}
%    \item \textit{waiting}: This mission has at least one super mission that is not finished. Missions with \textit{waiting} status is not allowed to start.
%    \item \textit{ready to compute}: This mission has no super mission that is not finished. Missions with \textit{ready to compute} status is allowed to start. They are requested to start once the manager expects that there are vacant memory and CPU cores.
%    \item \textit{computing}: Once a mission with status \textit{ready to compute} is arranged to start, its status changes to \textit{computing}.
%    \item \textit{finished}: Once a mission with status \textit{computing} finished running, its status changes to \textit{finished}.
%\end{itemize}
The task manager is in charge of the evolving of the registration table in the following ways,
%The mission statuses in the registration table are evolving in the following ways,
\begin{itemize}
    \item At the beginning, all mission statuses are initialized as \textit{waiting}.
    \item The manager reads the registration table periodically. 
    \item Once the manager detects that a \textit{waiting} mission has no unfinished super mission, it modifies its status to \textit{ready to compute}.
    \item Once the manager expects that there are available computation resources, it modifies the statuses of certain number of missions with \textit{ready to compute} status to \textit{computing} status, and, at the same time, send a message to workers to start the computing of those missions.
    \item Once a worker finished computing, it reports to the registration table to modify the status of the mission as \textit{finished}.
\end{itemize}

In {\sc NeatIBP}, we inserted a mission status monitor to show the contents of the registration table in real time, which describes status of all the missions. We will show this in Section \ref{subsec: manual running}.

\subsection{Generating IBP relations in one sector} \label{subsec:Generating IBP relations in one sector}
%In section \ref{subsec:Generating IBP relations referring to sectors}, we have introduced a parallel scheme to devide the IBP relation generation into missions that generates a fine set of IBP relations in one sector. 
In this section, the IBP generation in our program is introduced.
IBP relations can be generated via a seeding process introduced in Section~\ref{subsec:IBP relations from the Baikov representation} from the formal IBP relations derived in the Baikov representation. Note that unlike Laporta algorithm, given a fixed seed, IBPs from the module intersection or syzygy can have quite uneven numerator degrees. Therefore, for the given maximal numerator degree, we use different seeds for each formal IBP in eqn. \eqref{eq:Baikov IBP} to make sure that the resulting IBP relations have the same maximal numerator degree. We call this seeding process {\it Zurich seeding}\footnote{One of the authors of this paper used the process frequently when he was working in Zurich.}.

The IBP relations can be expressed as a linear system as follows
\begin{equation}\label{eq：specific IBP}
    0=\sum_{j} c^{\text{h}}_{ij} I^{\text{h}}_j+\sum_{k} c^{\text{t}}_{ik} I^{\text{t}}_k
\end{equation}
where $I^{\text{h}}_j$ and $I^{\text{t}}_k$ stands for head and tail integrals relevant to the IBP system generated from the current sector. %We require that the system is enough to reduce all target integrals of the current sector to its master integrals while considering the tail integrals as zero. Under this condition, we require the IBP relations to be linearly independent and minimum-necessary, considering the tail integrals as zero.

To make sure the IBP system can reduce the target integrals to master integrals in this sector and the tail integrals, row reduction operations on the matrix of the coefficients $c^{\text{h}}_{ij}$ would be applied. %By reducing the coefficient matrix to a row echelon form, we can determine which integrals are reduced. They correspond to the pivots of the reduced matrix. The other relevant integrals are considered to be not reduced by the given IBP system. 
For the row reduction, the ordering of the integrals in \eqref{eq：specific IBP} should be arranged by our preference. %In the row reduction method, the integrals in the front (with smaller subscripts) tend to become reduced integrals. Thus, we should put the complicated integrals, which are what we want to reduce, in the front. They are usually integrals with higher numerator of denominator degrees.

The master integrals of the current sector are also determined during the row reduction. On the other hand, We find that the foreknow information on the master integral can sometimes reduce the size of IBP seeds\footnote{Because we do not need to add more seeds to verify whether the set of master integrals is stable when numerator degree increases.}, and thus reduce the memory usage. For this purpose, we use Lee's critical counting \cite{Lee:2013hzt} and also the maximal cut syzygy method in {\sc Azurite} \cite{Georgoudis:2016wff}. Specifically, in {\sc NeatIBP} v1.0, we embed a minimal version of {\sc Azurite}, named as {\sc Azuritino}, to determine the master integrals as a running option. 

We remark that, we use Lee's critical point counting here as an ``order of magnitude'' estimation of the number of master integrals, to set up the initial seeding range for \textsc{Azuritino}.

%In our consideration of the first condition, it is important to determine the master integrals in the current sector. By definition, master integrals are those integrals who can never be reduced by any IBP systems. Together with some assumptions based on our experience, we can determine the master integrals using the Laporta algorithm. Firstly, we generate a certain number of large enough IBP system whose irreduced integrals are assumed to contain all master integrals. Thus, the intersection of the sets of irreduced integrals also contains all the master integrals. With the number of IBP system to be large enough, we expect that the intersection set is the set of master integrals.

By a standard linear algebra computation, a column reduction of $c^{\text{h}}_{ij}$ matrix selects independent IBP relations on the current sector. Here we prefer IBP relations with lower tail integral degrees in the selection. Schematically, after removing the linearly dependent relations, we have
\begin{equation}\label{eq：indep specific IBP}
    0=\sum_{j} \tilde c^{\text{h}}_{ij} I^{\text{h}}_j+\sum_{k} \tilde c^{\text{t}}_{ik} I^{\text{t}}_k
\end{equation}
Furthermore, from the linearly independent IBPs,
%Notice that in this process the IBP relations that corresponds to the rows with smaller row number $i$ tends to remain. Thus, we need to sort the IBP relations by our preference and put those we want to keep in the front. Our preference criterion varies, including number of integrals relevant and the complexity of the integrals. 
%One of the most important criteria is considering the [[[numerator degrees]]] of the tail integrals relevant to by the IBP relation. 
%As stated in sub section \ref{subsec:Generating IBP relations referring to sectors}, these tail integrals will appear as target integrals in the sub sectors, their numerator degree affects the complexity of the computations in the sub sectors much. Thus, we prefer to put the IBP relations with tail integrals that are with less numerator degrees in the front.
%After selecting independent IBP system in the sub space of the current sector, 
we can select IBP relations that are actually needed to reduce the target integrals. Specifically, considering a row reduction,
\begin{equation}
    R_{ik}=\sum_j L_{ij}\Tilde{c}^{\text{h}}_{jk},
\end{equation}
where the matrix $R$ is a row echelon form. %Each row with a pivot in $R$ corresponds to a reduced integral and its reduction form as
%\begin{equation}\label{eq:solution from row reduction}
 %   I_{p(i)}=-\sum_{j=p(i)+1}^n R_{ij}I_j,
%\end{equation}
%where $p(i)$ stands for the pivot of the $i$-th row of $R$. 
Suppose that the $i$-th row corresponds to an input integral, the set
\begin{equation}
    \{j|L_{ij}\neq 0\},
\end{equation}
corresponds to the indices of used IBPs to reduce this particular input integral. In this manner, we can further select used IBPs from the independent IBP set.

%the row numbers that actually contributes to the solutions in form of %\eqref{eq:solution from row reduction} of all target integrals are in the set
%\begin{equation}
 %   \bigcup_{i\in\big\{k|I_{p(k)}\in \{\text{target integrals}\}\big\}}\%{j|L_{ij}\neq 0\}.
%\end{equation}
%The rows with subscript that does not belong to the above set are not needed for reducing the target integrals. Thus, they can be deleted. So can the corresponding IBP relations. This operation will make the IBP system further simplified.

We remark that the above row reduction operations are just for IBP relation selections. Therefore, we use generic kinematic values and finite field methods to speed up these steps. 
However, this could be a bit risky if the numeric point is not ``generic" enough or prime number associated with the finite field hits a singular case in the linear operators. In {\sc NeatIBP} v1.0, we added a check process to check the IBP relations, with other numeric kinematic points. Multiple kinematic points and prime numbers, and a majority vote process may make the selection more stable. However, this approach would change the parallelization scheme and hence we leave it for the future versions of {\sc NeatIBP}. 

In this version of {\sc NeatIBP}, we implement the row reduction over finite field by using the open-source package {\sc SpaSM} \cite{spasm}. 

{\sc NeatIBP} allows input integrals with or without multiple propagators. For integrals with multiple propagators, {\sc NeatIBP} serves as an IBP generator which avoids higher propagator indices and reduces the IBP system size as well. With {\sc NeatIBP} v1.0, we suggest to run tasks for input integrals with or without multiple propagators separately. One may avoid a large number of seeds by doing this. %with both high degree numerators and multiple propagators.

%considering introducing more numeric points while generating IBP relations and use a majority voting technique to detect the numerical points that leads to wrong results.

%the matrix entries, which are original expressions with variables, are taking their numerical values on a given generic numeric point of the variables they rely on. Furthermore, the row reduction calculations are performed over finite fields. We apply such technique in order to avoid the complicated row reduction of matrices with rational function or rational number entries, which are with unbearable computational complexity. We are only interested in the pivots of the row echelon matrices; thus, a numerical-finite field method can be applied here. 

We remark that, in {\sc NeatIBP}, while solving module intersection problems in {\sc Singular}, a degree bound is frequently used. Once the total degree of an intermediate result reaches the given \texttt{degBound}, an early termination of Buchberger's algorithm %computation 
is triggered. This results in a faster computation, but possibly yields an incomplete basis, and thus incomplete generating system of the module intersection. We observe that there are cases, where the module intersection result obtained with degree bound, together with seeds without increased indices, may lead to a larger irreducible integral set than the set of master integrals. 
%IBP relations generated from seeds with denominator power indices strictly not increased could be not enough to reduce all targets to master integrals.
In these cases, {\sc NeatIBP} v1.0 can automatically search\footnote{We describe the strategy of how to search for them in a footnote in Section \ref{subsec: inputs} around the discussion on the option \texttt{StrictDenominatorPowerIndices}.} for certain seeds with increased denominator power indices, to further reduce the irreducible integral set. %Such seeds are selected to make sure their quantity and degrees are minimum-necessary. 
This mechanism is designed for complicated diagrams where we have to use low degree bounds in {\sc Singular} to make the module intersection computation less time-consuming.

\subsection{Additional remarks}
The whole process of {\sc NeatIBP} is in three steps, the initialization, the parallel missions between sectors and the summary.

In the initialization step, we determine all sectors that are relevant to the missions according to the target integrals the user input. Among the sectors, there are some zero sectors. We treat all target integrals in these zero sectors as $0$ in the whole process.

{\sc NeatIBP} allows the user to choose whether the symmetry is needed. If so, we need to consider both symmetries inside sectors and between sectors. The symmetries between sectors are found in the initialization step. Through this, some nonzero sectors\footnote{In this context, they are called the \textit{source sector} of its image sector.} can be mapped to its \textit{image sector}. We select a set of nonzero sectors, called \textit{unique sectors}, such that all other nonzero sectors can be mapped to a unique sector, and a unique sector cannot be mapped to another unique sector. {\sc NeatIBP} converts all integrals in the non-unique sectors to their images after symmetry mapping, which are combinations of integrals from unique sectors.

%The second step of {\sc NeatIBP} is to generate IBP relations in parallel. There are some remarks about the symmetries.

In the second step, the symmetries inside a sector can be derived using methods stated in Subsection \ref{subsec: symmetry review}. %By a similar seeding process, we can get specific symmetry relations. 
We treat these symmetry relations in the same way as IBP relations and put them together to form a linear system after seeding. %The non-unique sector integrals will be mapped to integrals in unique sectors. Therefore, the IBP seeding would be carried out only in unique sectors. %Thus, the concept of sub sectors in sub section \ref{subsec:Generating IBP relations referring to sectors} is generalized when considering symmetries, in order to control the dependency between different missions. 

%\subsection{Generating IBP relations in one sector}
%In this subsection, we will introduce how do we generate the minimum needed IBP relations in order to reduce the target integrals, which are integrals to be reduced. From the Baikov representation and the module intersection method, we are able to get some formal IBP relations. After seeding, we get many specific IBP relations. Among the specific relations, many are not linearly independent to others or useless for reducing the targets. We need to choose those that are really needed and independent. 

%In the above steps, we need to perform row reduce among matrices. In order to make it faster, we will take the kinematic variables, such as masses and Mandelstam variables, as rational numbers, to get numeric IBP relations. Then we are able to perform row reduce of the numeric matrices on finite field. This will make it much {\red (much?)} faster.

%In the next parts of this sub section, we are going to show some details about the algorithms of the above steps.

\section{Manual} \label{sec:manual}
In this section, we provide a manual describing the installation and usage of {\sc NeatIBP}. Note that this manual is based on the most recent version (v1.0.2.4) at the time that this paper is finished. This version is complete in the sense that it implements the basic functions introduced in the previous sections. {\sc NeatIBP} will be continually updated with further improvement. For this, users can refer to the {\sc NeatIBP} GitHub repository. %mentioned in Section \ref{subsec:installation} below to follow the updates.

\subsection{Installation}\label{subsec:installation}
{\sc NeatIBP} is a {\sc Mathematica}/{\sc Singular} package. 
One can obtain {\sc NeatIBP} from the following GitHub repository link, which also provides installation instructions.\smallskip

\centerline{\href{https://github.com/yzhphy/NeatIBP}{https://github.com/yzhphy/NeatIBP}}\smallskip

\noindent The user can install {\sc NeatIBP} by %going to the installation path, and running:
\begin{lstlisting}[language=bash]
git clone https://github.com/yzhphy/NeatIBP.git
\end{lstlisting}
following the \texttt{Readme.md} document of the repository. The dependencies of {\sc NeatIBP} are {\sc Singular} \cite{DGPS} and {\sc SpaSM} \cite{spasm}, so these have to be installed before usage. If {\sc SpaSM} is not present on the machine yet, we recommend to install it using or along the lines of script provided in the {\sc NeatIBP} repository. 

As an alternative way of installation, we also provide a package for the supercomputing package manager \textsc{Spack} \cite{7832814}, which installs {\sc NeatIBP} along with all of its dependencies. Please follow the respective section of the \texttt{Readme.md} document. 

\subsection{Preparing inputs}\label{subsec: inputs}
To utilize {\sc NeatIBP} for a Feynman integrals reduction task, the following three files are required in the working directory in order to run the program:
\begin{enumerate}
\item A configuration file named ``\texttt{config.txt}" containing the necessary settings. 
\item A kinematics file, which by default is named ``\texttt{kinematics.txt}", containing the kinematic information for the Feynman diagram.
\item A target integral file, which by default is named ``\texttt{targetIntegrals.txt}", containing the target integrals which are supposed to be reduced.
\end{enumerate}

The input files should be prepared in form of Mathematica readable files using the command \texttt{Get}. In the configuration file, settings that control the execution of {\sc NeatIBP} are specified. %Note that in the configuration file does not need to include all available settings. 
If an option value is not given, the program refers to default values given in \texttt{NeatIBP/default\_settings.txt}. We list some of the settings which are frequently used:
\begin{enumerate}
    \item \texttt{kinematicsFile} and \texttt{targetIntegralsFile}: These variables contain the names of the txt files which specify the kinematics of the Feynman diagram and the target integrals, respectively. Please make sure that the file names are given with absolute paths. To do so, one can use \texttt{workingPath} which points to the working directory (the directory of the configuration file). 
    \item \texttt{SingularApp}: This variables tells {\sc NeatIBP} how to run Singular. Its value is a string with the command running {\sc Singular} in the shell.
    \item \texttt{SparseRREF\`{ }SpaSMLibrary}: This variable tells {\sc NeatIBP} where to find \textsc{SpaSM}. Its value is a string containing the absolute path to the file ``\texttt{libspasm.so}''.
    \item \texttt{ReductionOutputName} and \texttt{outputPath}: If \texttt{outputPath} is set to the value \texttt{Automatic} (default value), the output of {\sc NeatIBP} will be written to the subfolder with name \texttt{ReductionOutputName} inside the subfolder \texttt{outputs} in \texttt{workingPath}. If the output should be written to a different location, \texttt{outputPath} can be set.%to be the absolute path of this location
    \footnote{If \texttt{outputPath} is set to \texttt{Automatic} and the output path already exists, then {\sc NeatIBP} will ask the user whether to overwrite the contents. If the folder is user-specified, it is the user's responsibility to delete data from previous runs.} %{\sc NeatIBP} will not delete a folder which is not an Automatic output folder} %considering security.}.

    \item \texttt{OptionSimplification}: This is an integer value selecting the simplification mode used to simplify the results of the module intersection in \textsc{Singular}. The simplification mode is explained in the \textsc{Singular} documentation, see
    \begin{center}
        \url{https://www.singular.uni-kl.de/Manual/4-3-2/sing_349.htm}.
    \end{center}
     
     \item \texttt{FiniteFieldModulus}: The finite field modulus used in {\sc Spasm} for NeatIBP. This value should be a prime number not larger than $46337$.
     \item \texttt{AzuritinoIntersectionDegreeBound} specifies the degree bound for module intersection computations via {\sc Singular} in {\sc Azuritino}, while  \\ \texttt{NeatIBPIntersectionDegreeBound} gives the degree bound for the formal IBP generation step in {\sc NeatIBP}. For more details on how to specify degree bounds, please refer to 
     \begin{center}
         \url{https://www.singular.uni-kl.de/Manual/4-3-2/sing_399.htm}.
     \end{center}

    \item \texttt{IntegralOrder}: A variable controlling the ordering of Feynamn integrals. It can take the values \texttt{MultiplePropagatorElimination}, \\ \texttt{ISPElimination}, or \texttt{Global}. The three options correspond to different preferences of orderring, respectively reducing the degree of denominators, numerators, or both (the total absolute degree), as priority. Note that, when \textsc{Azuritino} is enabled (which is by default), since it finds master integrals without double propagators, \texttt{IntegralOrder} must be set to \texttt{MultiplePropagatorElimination}.
    \item \texttt{NeedSymmetry}: A Boolean type variable, which determines whether symmetries of Feynman integrals are considered.
    \item \texttt{CutIndices}: A list containing the indices of the cut propagators. 
    %\item \texttt{GenericD}: The replacement rule of a generic choice of a numerical (rational number) value of spacetime dimension $d$. This variable can also be defined in the kinematics file. If so, the value defined in the configuration file will be override.
    %\item \texttt{seedingViaFIBPFunction}: A Boole value to determine whether to use abstract functions while seeding. According to our observation, this will make the seeding faster, although it needs an additional step to derive the abstract functions in Mathematica.
    %\item \texttt{SowAndReap}: A Boole value to determine whether we should use {\sc Mathematica} Sow-and-Reap implementation in Zurich seeding in the reduction process.
    %\item \texttt{ParallelInFindingSectorMaps}: A Boole value to determine whether we should parallelize the finding of symmetry maps between sectors.
    \item \texttt{SeedingAdditionalDegree}: The maximum of the additional numerator degree considered relative to the degree of the target integrals. 
    \item \texttt{MIFromAzuritino}: A Boolean type variable, which determines whether to use {\sc Azuritino} to find master integrals.% {\sc Azuritino} is a minimal version of {\sc Azurite} \cite{Georgoudis:2016wff}.
    \item \texttt{CriticalPointInAzuritino}: A Boolean type variable determining whether to use Lee's critical point counting method \cite{Lee:2013hzt} to estimate the amount of master integrals before using {\sc Azuritino}.
    \item \texttt{StrictMI}: If this  Boolean type variable is set to \texttt{True}, then {\sc NeatIBP} terminates with an errror if the IBP relations generated are not sufficient to reduce the target integrals to the pre-determined master integrals. Otherwise, {\sc NeatIBP} uses a (typically slightly) redundant integral basis. 
    \item %\texttt{StrictDenominatorPowerIndices} %and \\ %\texttt{AllowedDenominatorPowerLift}: 
    If \texttt{StrictDenominatorPowerIndices} is \texttt{False}, {\sc NeatIBP} will try to introduce a minimally necessary set of seeds with increased indices for denominators, in case the generated IBP system cannot reduce the target to master integrals\footnote{This function is called \texttt{AdditionalIBPs} in \textsc{NeatIBP}. It firstly tries to add seeds with only one denominator index increased at each position and test if these seeds generate a complete linear system. Otherwise, according to the value of \texttt{AllowedDenominatorPowerLift}, the function adds seeds with multiple increased indices for denominators. For example, if \texttt{AllowedDenominatorPowerLift} is set to $2$, there may be index increasement of type $(+1,+1,0,0,\cdots)$ or $(+2,0,0,0,\cdots)$.}. %, when the additional seed numerator reaches \texttt{SeedingAdditionalDegree}. 
    
    \texttt{AllowedDenominatorPowerLift} is the allowed total denominator degree increase in the additional seeds. Note that setting the variable \texttt{AllowedDenominatorPowerLift} to $0$ is equivalent to setting  the variable \texttt{StrictDenominatorPowerIndices} to \texttt{True}.
    \item \texttt{MemoryUsedLimit}: The limit for memory usage specified in megabytes. {\sc NeatIBP} will try to estimate the memory usage of newly scheduled parallel tasks, and try to control (but not to guarantee) the memory usage below this limit.
    \item \texttt{ThreadUsedLimit}: The maximum number of {\sc NeatIBP} parallel computation. This value is used in two circumstances. First, when finding symmetries between sectors, \textsc{NeatIBP} by default uses \texttt{ParallelTable}. This value is the kernel it launches using \texttt{LaunchKernels} (if it is not larger than \texttt{\$ProcessorCount}). Second, in the IBP relation generation step, this value is tha maximum number of sectors that are synchronously computed. Note that, in this step, the task manager also needs to occupy a \textsc{Mathematica} kernel. So, the maximum kernel usage in this step is \texttt{ThreadUsedLimit}+1. Also, if you turn on a monitor, since it also occupies a kernel, the number of kernels you may use is \texttt{ThreadUsedLimit}+2.
    \item Using the Boolean type variables \texttt{DeleteSingularScriptFiles} and \texttt{DeleteSingularResultFiles}, one can determine whether \textsc{NeatIBP} deletes on completion of the computation the temporary {\sc Singular} input and output files, respectively.
    
\end{enumerate}
The user can also request in the configuration file various additional outputs. For details, see the section ``additional outputs'' in the default configuration file
\texttt{NeatIBP/default\_settings.txt}.\bigskip

\noindent In the kinematics file, the user shall provide the following variables:
\begin{enumerate}
    \item \texttt{LoopMomenta}: A list for the loop momenta of the Feynman diagram.
    \item \texttt{ExternalMomenta}: A list containing all external momenta of the Feynman diagram.
    \item \texttt{Propagators}: A list with the propagators of the Feynman diagram. 
    \item \texttt{Kinematics}: A list containing replacement rules for the kinematic conditions of the external momenta.
    \item \texttt{GenericPoint}: A list containing replacement rules corresponding to a general numeric point (with rational coordinates) for scalar kinematic variables appearing in \texttt{Kinematics} and \texttt{Propagators}.
    \item \texttt{GenericD}: A replacement rule corresponding to a general numeric value (rational number) for the spacetime dimension $d$.
    
\end{enumerate}

%In the kinematics file, all the variables should be set in order to run {\sc NeatIBP}.
\noindent In the target integral file, the user shall prepare a list of the target integrals. Here, the integral $G_{\alpha_1,\cdots,\alpha_n}$ is expressed as \texttt{G[}${\alpha_1},\cdots,\alpha_n$\texttt{]}.
\subsection{Running {\sc NeatIBP}}\label{subsec: manual running}
After preparing the above-mentioned input files in the working directory, the user can execute {\sc NeatIBP} to generate the IBP relations. Assuming the user has installed {\sc NeatIBP} in the directory \texttt{/SomePath/NeatIBP/}, it can be run by changing directory to the working path and executing the command
\begin{lstlisting}
/SomePath/NeatIBP/run.sh
\end{lstlisting}

While {\sc NeatIBP} is running, the user can execute in a separate terminal window in the working path the command
\begin{lstlisting}
/SomePath/NeatIBP/monitor.sh
\end{lstlisting}
to launch a monitor which shows the current status of the parallel missions. 

An example of a monitor output is
\begin{lstlisting}[breaklines=false,columns=fullflexible]
----------------------------------------------
2023.4.10 23:17:53
42 sector(s) waiting super sector(s): 244,241,236,233,229,205...
0 sector(s) ready to compute.
14 sector(s) computing: 245,237,218,217,211,206...
81 sector(s) finished: 255,254,253,251,247,239...
----------------------------------------------
\end{lstlisting}
In the output, the sector numbers are the binary ids of the sectors.

Note that, in case a worker process terminates unexpectedly during computation, this will not be reported to the registration table and the management system will consider the job to still be computing. To recognize this, the monitor has a process detection, which will report in case of an unexpected termination of a process a warning of the form
\begin{lstlisting}[breaklines=false,columns=fullflexible]
----------------------------------------------
2023.3.17 15:32:25
128 sector(s) waiting super sector(s): 252,250,249,246,245,243...
0 sector(s) ready to compute.
7 sector(s) computing: 254,253,251,247,239,223...
2 sector(s) finished: 255,127.
******** 
Warning:
1 sector(s) lost: 191.
The corresponding process(es) cannot be detected. Maybe they terminated 
unexpectedly.
----------------------------------------------
\end{lstlisting}
In the above example, the computing process of sector 191 was terminated manually while computing, resulting in the warning.

{\sc NeatIBP} supports to resume computation if a process is terminated by hand or unexpectedly. To do this, make sure the execution of {\sc NeatIBP} in the respective working path is stopped, then run the following command in the working path
\begin{lstlisting}
/SomePath/NeatIBP/continue.sh
\end{lstlisting}
{\sc NeatIBP} will restart the sectors that have not been finished yet.

In the Github repository of {\sc NeatIBP}, we provide some examples for how to use {\sc NeatIBP}. In the current version, one of the examples (dbox) include the output results. User can refer to this manual and try the examples by themselves. Note that, in the examples, we provided some convenient scripts to apply the operations, which again are called \texttt{run.sh}, \texttt{monitor.sh}, and \texttt{continue.sh}. User can modify the file \texttt{packagePath.txt} such that it contains the absolute path where \textsc{NeatIBP} was installed. Then, open a terminal in these example working folders, and use
\begin{lstlisting}
./run.sh
\end{lstlisting}
to run {\sc NeatIBP}, use
\begin{lstlisting}
./monitor.sh
\end{lstlisting}
to turn on a monitor, or use
\begin{lstlisting}
./continue.sh
\end{lstlisting}
to continue an unfinished computation. For computation in other directories, the user can copy the file \texttt{packagePath.txt} along with respective three scripts from the example's directory to the working directory and them do the same.

\subsection{Output generated}
After {\sc NeatIBP} has finished running, its output will be exported into three sub directories of the output path, namely \texttt{inputs}, \texttt{results}, and \texttt{tmp}. The \texttt{inputs} folder will contain the three input files described in Subsection~\ref{subsec: inputs}. The \texttt{results} folder will contain four files, and a minimum of two sub folders. The file \texttt{IBP\_all.txt} lists all IBP relations generated by {\sc NeatIBP}. The file \texttt{MI\_all.txt} lists all master integrals. The file \texttt{OrderedIntegrals.txt} lists all integrals occurring in the IBP relations, in an integral ordering from high to low. The file \texttt{summary.txt} is a log file containing a summary of the computation. The sub folders \texttt{results/IBP/} and \texttt{results/MI/} contain txt files with IBP relations and master integrals on each sector, respectively. If the user requested additional outputs in the configuration file, there will be sub folder \texttt{additional\_outputs} containing the requested additional outputs.
The folder \texttt{tmp}, contains some temporary files created during runtime. %Among them, there are some useful log files that record the detailed information of each mission.

As discussed in Subsection \ref{subsec:Generating IBP relations in one sector}, the finite field row reduction in {\sc NeatIBP} using a numeric point may hit singularities. Thus, after {\sc NeatIBP} finished running, it is recommended to check the results on further different numeric points. In {\sc NeatIBP}, we provide an automatic checker to test if the IBP relations generated can reduce the target integrals to master integrals. To do this, open a terminal in any folder and run 
\begin{lstlisting}
math -script /SomePath/NeatIBP/CheckIBP.wl [path]
\end{lstlisting}
where \texttt{[path]} is the output path of the {\sc NeatIBP} results. The program will perform a row reduction on a different random numeric point to see if the IBP system is sufficient to reduce the target integrals. It is recommended to execute the check several times to sufficiently reduce the probability of a mistake.

\subsection{Row reduction using {\sc SpaSM}}

As mentioned before, this version of {\sc NeatIBP} relies on {\sc SpaSM} \cite{spasm} for row reduction over finite fields. To connect {\sc SpaSM} to {\sc Mathematica}, we use {\sc LibraryLink}, a component of {\sc Mathematica}, which can connect to C libraries and supply an interface. For communication, specific data structures are used in our C code, which then is packaged as a library called \texttt{SparseRREF.so}, and an interface file called \texttt{SparseRREF.m}. %is made to introduce such library into {\sc Mathematica}. 
%
%{\sc SpaSM} is a C library devoted to do such sparse RREF (Row Reduction Echelon Form) modulo a small prime \textit{p}. The core of the library is an implementation of the GPLU algorithm. 
%
%{\sc SpaSM} provides proper data structure of sparse integer matrices and functions for matrix operations, including the essential \textit{spasm\_LU} for LU decomposition and \textit{spasm\_sparse\_forward\_solve} for solving sparse linear system. We implement the RREF procedure with these functions in C language. 
%
In {\sc NeatIBP}, this interface to {\sc SpaSM} is included and automatically used. 

The interface component can also be used individually for other purposes. To do this, one can run the following command in {\sc Mathematica}

\begin{lstlisting}[language=Mathematica]
Get["SomePath/NeatIBP/SparseRREF/SparseRREF.m"]
\end{lstlisting}
and use the command
\begin{lstlisting}[language=Mathematica]
SRSparseRowReduce[M, Modulus->p]
\end{lstlisting}
to obtain the RREF (reduced row echelon form) from an sparse input matrix \textit{M} (which should be a \texttt{SparseArray}) given over the rational numbers which is reduced to the finite prime field $\mathbb{F}_p$ modulo the prime \textit{p} in case of co-prime denominators. The output will be a sparse echelon-form modulo \textit{p} of {\sc Mathematica} type \texttt{SparseArray}.

\section{Examples} \label{sec:examples}
In this section, we provide examples to illustrate the performance of {\sc NeatIBP}. This section gives details on three different examples: a 2-loop 5-point massless planar diagram, a 2-loop 4-point non-planar diagram with internal and external massive lines, and a 3-loop 4-point planar diagram with one external massive line. The input files of these examples can be found in the Github repository, in folder ``examples/examples\_in\_the\_papers/''. The input and output files can be found in the Dropbox links shown later.

\subsection{A pentagon-box diagram example}
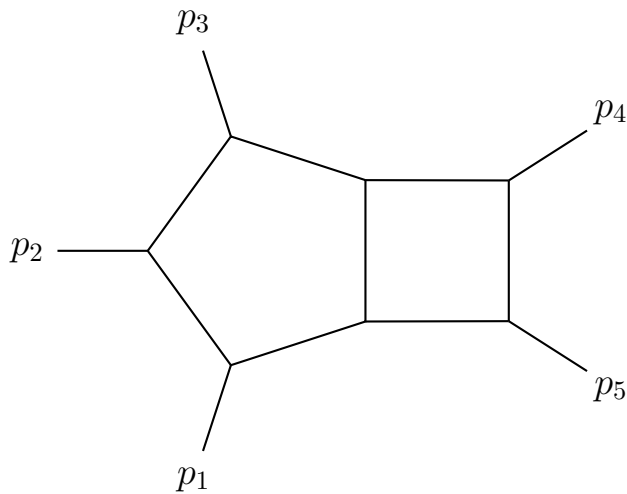
\begin{figure}[htpb]
\centering
\hspace{-1.8cm}
\begin{tikzpicture}[scale=0.8]
\newcommand{\sidelength}{2}
\draw[thick] (36:\sidelength)
\foreach \x in {36,108,180,252,324} {
--(\x:\sidelength)} --cycle;
\draw[thick](36:2)--(4,1.17);
\draw[thick](324:2)--(4,-1.17);
\draw[thick](4,1.17)--(4,-1.17);
\draw[thick](108:2)--(108:3.5);
\draw[thick](180:2)--(-3.5,0);
\draw[thick](252:2)--(252:3.5);
\draw[thick](4,-1.17)--(5.3,-2);
\draw[thick](4,1.17)--(5.3,2);

\node[font=\large\bfseries] at (252:4){$p_1$};
\node[font=\large\bfseries] at (-4,0){$p_2$};
\node[font=\large\bfseries] at (108:4){$p_3$};
\node[font=\large\bfseries] at (5.7,2.3){$p_4$};
\node[font=\large\bfseries] at (5.7,-2.3){$p_5$};
\end{tikzpicture}
\caption{A two-loop five-point diagram example}
\label{fig:pentabox}
\end{figure}

In this subsection, we consider the two-loop five-point diagram as depicted %We use {\sc NeatIBP} to generate a small sized IBP system enough to reduce certain given target integrals of this diagram. % We also compare the sizes of the IBP system generated by {\sc NeatIBP} with those generated during a corresponding reduction task in FIRE6 \cite{Smirnov:2019qkx}, which uses traditional Laporta's algorithm, to see by how many the size decreased.
 in Figure~\ref{fig:pentabox}. 
The paper \cite{Gluza:2010ws} which originally introduced the syzygy IBP method   used this diagram as an example, but did not cover all the sub sectors.  

The propagators of this diagram are
\begin{equation}
\begin{aligned}
&D_1=l_1^2,\quad
D_2=(l_1-p_1)^2,\quad
D_3=(l_1-p_{12})^2,\quad
D_4=(l_1-p_{123})^2,\nonumber\\
&D_5=l_2^2,\quad
D_6=(l_2+p_{1234})^2,\quad
D_7=(l_2+p_{123})^2, \quad
D_8=(l_1+l_2)^2,\nonumber\\
&D_9=(l_1-p_{1234})^2,\quad
D_{10}=(l_2+p_1)^2,\quad
D_{11}=(l_2+p_2)^2, 
\end{aligned}
\end{equation}
where $p_{i\cdots j}:=p_i+\cdots+p_j$. The last three propagators are irreducible scalar products (ISPs). The external kinematic conditions are
\begin{equation}
\begin{aligned}
&p_1^2 = p_2^2 = p_3^2 =  p_4^2 = 0,\quad
p_1\cdot p_2 = \frac{s_{12}}{2}, \nonumber\\&
  p_1 \cdot p_3 =  \frac{-s_{12} - s_{23} + s_{45}}{2}, \quad
  p_1\cdot p_4 = \frac{-s_{15} + s_{23} - s_{45}}{2}, \nonumber\\&
  p_2\cdot p_3 = \frac{s_{23}}{2}, \quad
  p_2\cdot p_4 = \frac{s_{15} - s_{23} - s_{34}}{2}, \quad
  p_3\cdot p_4 = \frac{s_{34}}{2}
\end{aligned}
\end{equation}
The momentum conservation condition in this diagram is $p_1+p_2+p_3+p_4+p_5=0.$

We demonstrate this example using two sets of target integrals as inputs. The first set of target integrals is to show the performance on numerator reduction. It consists of $2483$ randomly chosen integrals from sector $\{1, 2, 3, 4, 5, 6, 7, 8\}$ and its subsectors. They have maximum numerator degree $5$, and are free of multiple denominators. Using {\sc NeatIBP} (v1.0.2.3), we find $61$ master integrals and get an IBP system with $14120$ IBP identities, which are enough to reduce all the target integrals to master integrals.

%The IBP identities contains $14262$ related integrals. 

The above computation was carried out using $10$ CPU cores and $128$GB RAM. The computation took about $27$ minutes. The input and some output files can be obtained at the following link:
\begin{quote}
\url{https://www.dropbox.com/s/rrwillc0xyv30pf/pb.tar.gz}
\end{quote}
As a reference, {\sc FIRE6} \cite{Smirnov:2019qkx} used $11207942$ IBP identities in the reduction of the same target integrals. In the computation, we ran {\sc FIRE6} with kinematic variables set to be rational numbers. 

The second set of targets contains $880$ integrals, which come from the derivatives of the $61$ master integrals. This set shows the performance of {\sc NeatIBP} for integrals with multiple propagators. By generating $3313$ IBP identities, {\sc NeatIBP} relates the given $880$ integrals to the $61$ master integrals. The computation finishes in $17$ minutes on $10$ CPU cores and $128$GB RAM. As a reference, {\sc FIRE6} reduced those target integrals using $1010236$ IBP identities. The input and relevant output files can be downloaded at the following link:
\begin{quote}
\url{https://www.dropbox.com/s/vhzilt837nhnll8/pb_D.tar.gz}
\end{quote}
\subsection{A non-planar two-loop massive diagram example}
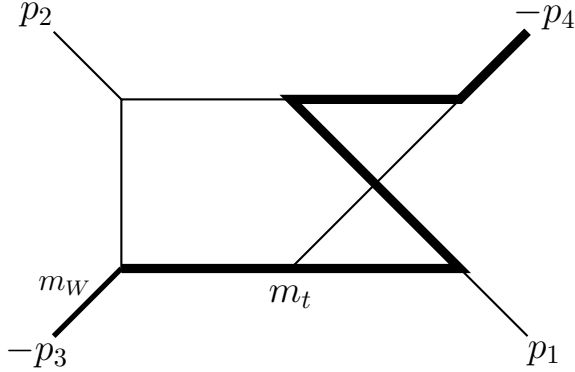
\begin{figure}[hbtp]
\begin{center}
\begin{tikzpicture}[scale=0.75]
\draw[line width = 2pt](-1.2,-1.2)--(0,0);
\draw[thick](-1.2,4.2)--(0,3);
\draw[thick](7.2,-1.2)--(6,0);
\draw[line width = 3.5pt](7.2,4.2)--(6,3)--(3,3)--(6,0)--(0,0);
\draw[thick](0,0)--(0,3)--(3,3);
\draw[thick](3,0)--(6,3);

%\draw(0,0)--(6,0);
%\draw(0,3)--(6,3);

%\draw (6,3) -- (3,0);
%\draw (3,3) -- (6,0);
%\draw (0,3) -- (0,0);

\node[font=\large\bfseries]  at (-1.5,-1.5){$-p_3$};
\node[font=\large\bfseries]  at (-1.5,4.5){$p_2$};
\node[font=\large\bfseries]  at (7.5,4.5){$-p_4$};
\node[font=\large\bfseries]  at (7.5,-1.5){$p_1$};

\node[font=\large\bfseries]  at (3,-0.5){$m_t$};
\node[font=\bfseries]  at (-1,-0.3){$m_W$};

\end{tikzpicture}
\caption{A nonplanar two-loop diagram for the tW production}
\label{fig:The non-planar 2-loop 4-point diagram}
\end{center}
\end{figure}

In this subsection, we discuss an example from the top quark phenomenology. The diagram is given in Figure \ref{fig:The non-planar 2-loop 4-point diagram}. This nonplanar diagram is one of the most complicated two-loop diagrams contributing to the NNLO corrections for $tW$ production at hadron colliders in~\cite{Chen:2022pdw} (and the references therein). 

The propagators of this diagram are 
\begin{equation}
\begin{aligned}
&D_1=-m_t^2 + l_1^2,\quad
D_2= -m_t^2 + l_2^2, \quad
D_3=(l_1 + l_2)^2,\nonumber\\&
D_4= -m_t^2 + (l_1+p_1 )^2, \quad
D_5=-m_t^2 + (l_1 + l_2+p_1 + p_2 - p_3 )^2, \nonumber\\&
D_6=(l_2+p_2 - p_3 )^2,\quad
D_7=( l_2-p_3 )^2, \quad
D_8= (l_1-p_1 + p_3 )^2, \nonumber\\&
D_9= ( l_2-p_1 - p_2 + p_3 )^2.
\end{aligned}
\end{equation}
Its external kinematic conditions are
\begin{equation}
\begin{aligned}
&p_1^2=p_2^2=0,\quad p_3^2=m_W^2 ,\quad
p_1\cdot p_2 = \frac{m_t^2 + m_W^2 - t - u}{2},\nonumber\\&
p_1\cdot p_3 = \frac{m_W^2 - t}{2},\quad
p_2\cdot p_3 = \frac{m_W^2 - u}{2}
\end{aligned}
\end{equation}
The momentum conservation condition in this example is $ p_1+p_2=p_3+p_4$.

The $597$ target integrals for the two-loop scattering amplitudes $pp \to tW$, are from sector $\{1,2,3,4,5,6,7\}$ and its subsectors.  Using {\sc NeatIBP} (v1.0.2.3), we generated $7169$ IBP relations which are sufficient to reduce all the target integrals to $90$ master integrals. %The IBP relations contains 7424 related integrals. 

The computation was performed using $20$ CPU cores and $128$GB RAM. The computation took about $1.5$ hours.  The input and relevant output files can be obtained at the following link:
\begin{quote}
\hspace{-3mm}\url{https://www.dropbox.com/s/6wnfwfn0he47ztk/2l4pNP7.tar.gz}
\end{quote}

We remark that for this diagram the syzygy computation is rather expensive. Using the degree bound $5$ for the computation in {\sc Singular}, we observe that we do not obtain a sufficiently large set of syzygies, and hence the number of resulting irreducible integrals is slightly larger than the expected $90$. Increasing the degree bound to $6$ makes the syzygy computation significantly more expensive. {\sc NeatIBP} then obtains a sufficient set of IBPs and finds $90$ master integrals by automatically seeding on some extra integrals with double propagators.

\subsection{A three-loop example}
\begin{figure}[hbtp]
\begin{center}
\begin{tikzpicture}[scale=0.71]
\draw[thick] (0,0) rectangle (9,3);
\draw[thick] (3,0) -- (3,3);
\draw[thick] (6,0) -- (6,3);
\draw[thick](-1.5,-1.5)--(0,0);
\draw[thick](-1.5,4.5)--(0,3);
\draw[line width=3pt](10.5,-1.5)--(9,0);
\draw[thick](10.5,4.5)--(9,3);
\node[font=\large\bfseries] at (-1.8,-1.8){$p_1$};
\node[font=\large\bfseries] at (-1.8,4.8){$p_2$};
\node[font=\large\bfseries] at (10.8,4.8){$p_3$};
\node[font=\large\bfseries] at (10.8,-1.8){$p_4$};
\end{tikzpicture}
\caption{The planar triple box diagram with an external massive leg}
\label{fig:The planar triple box diagram}
\end{center}
\end{figure}
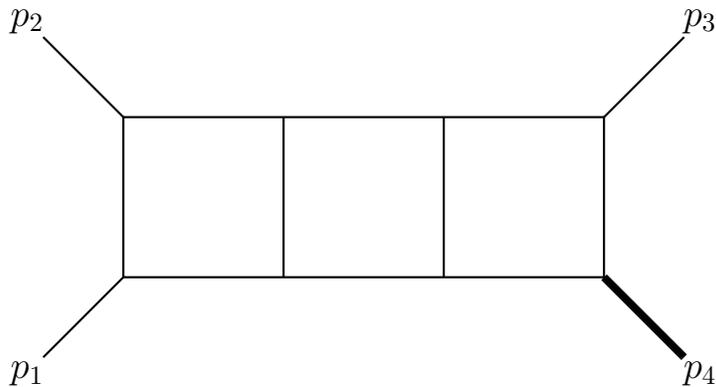

In this subsection, we present a three-loop four point example shown in Figure \ref{fig:The planar triple box diagram}.
Note that master integrals in related families are calculated in \cite{Henn:2023vbd}.

The propagators of this diagram are 
\begin{equation}
\begin{aligned}
&D_1=l_1^2,\quad
D_2=(l_1-p_1)^2,\quad
D_3=(l_1-p_{12})^2,\quad
D_4=l_3^2,\nonumber\\
&D_5=(l_3+p_{12})^2,\quad
D_6=(l_1+l_3)^2,\quad
D_7=(l_2-l_3)^2, \quad
D_8=l_2^2,\nonumber\\
&D_9=(l_2-l_4)^2,\quad
D_{10}=(l_2+p_{12})^2,\quad
D_{11}=(l_1+p_4)^2, \quad
D_{12}=(l_2+p_1)^2, \nonumber\\
&D_{13}=(l_3+p_1)^2, \quad
D_{14}=(l_3+p_4)^2, \quad
D_{15}=(l_1+l_2)^2.
\end{aligned}
\end{equation}
The external kinematic conditions are
\begin{equation}
\begin{aligned}
&p_1^2=p_2^2=0,\quad p_4^2=m^2 ,\quad
p_1\cdot p_2 = \frac{s}{2},\nonumber\\&
p_2\cdot p_4 = -\frac{s + t}{2},\quad
p_1\cdot p_4 = \frac{t-m^2}{2}.
\end{aligned}
\end{equation}
The momentum conservation condition for this diagram equals $p_1+p_2+p_3+p_4=0$.

We randomly pick $21185$ integrals as our target. They are in the sector
$\{1,2,3,4,5,6,7,8,9,10\}$ and its subsectors. The most complicated integral has the numerator degree $6$ in the top sector. Using {\sc NeatIBP} (v1.0.2.3), we found $83$ master integrals and acquired $200074$ IBP relations, which are sufficient to reduce all given target integrals to master integrals.
%The IBP relations contains $200928$ related integrals. 

This computation was carried out on a computer with 50 CPU cores and 1.5TB RAM and took about $6$ hours. The input and relevant output files can be downloaded at the link:
\begin{quote}
\url{https://www.dropbox.com/s/8mabbgauxpmxzrb/3l4p1m.tar.gz}
\end{quote}

\section{Summary} \label{sec:summary}

In this paper, we present the package {\sc NeatIBP} v1.0, for generating an IBP system which are sufficient to reduce the input target Feynman integrals to master integrals. The module intersection method, which is a contemporary variant of the original syzygy approach \cite{Gluza:2010ws}, is used to restrict the growth of propagator indices in the IBP generation. We expect that the size of this IBP system is of orders of magnitude smaller than that obtained from Laporta's algorithm. The resulting IBP system can be used for either finite field reduction with subsequent reconstruction of physical quantities, or for analytic reduction.

In this version of {\sc NeatIBP}, the module intersection computation is powered by the open-source computer algebra system {\sc Singular}, while IBP generation, workflow control and the I/O operations are controlled by {\sc Mathematica}. The selection of linearly independent and sufficient IBP relations is carried out using the C-library {\sc SpaSM}.  {\sc NeatIBP} v1.0 incorporates parallelization across different Feynman integral sectors. The execution is controlled by {\sc Mathematica} code, and the status of the computation is recorded in a registration table on disk. The status can be monitored by the user in the terminal. In the event that the execution of {\sc NeatIBP} is interrupted, it is possible to resume the computation using the saved temporary files stored on disk.

We anticipate the following updates in the future versions:
\begin{enumerate}
    \item Migrate the code to rely only on open-source software.
    \item Introduce advanced massive parallelization. This includes the introduction of an additional layer of parallelization within a single sector for concurrent execution of the seeding process.
    Further parallelization could arise from the different cuts of one Feynman integral family. A promising approach to effectively manage and intertwine  the multiple levels of parallelism is to realize the workflow management through the open source \textsc{Singular}/\textsc{GPI-Space} framework \cite{singgpi, BFK21}, which relies on the workflow management system {\sc GPI-Space} \cite{GSPC}, incorporates the computer algebra system \textsc{Singular}, which is used by \textsc{NeatIBP} for module intersection computations, and offers convenient installation using the \textsc{Spack} package manager \cite{spack}.
    
    \item Incorporate alternative approaches for module intersection or syzygy computation. In the current version of {\sc NeatIBP}, we apply the available general functionality of {\sc Singular} for computing module intersections. In future versions, we might consider to customize this implementation for our specific setting. Moreover, further methods, like linear algebra techniques \cite{Schabinger:2011dz} and the dual conformal symmetry method for syzygy generation \cite{Bern:2017gdk} will be considered in future versions.
    \item It is well known that the syzygy/module intersection method is very effective on cuts. {\sc NeatIBP} v1.0 can easily generate IBP relations on cuts. Future versions will find all spanning cuts automatically and build the rules for combining the results for the cuts.
\end{enumerate}

%\section{Raw Materials}

%{\red I am putting raw materials writing this paper here....}

%\begin{figure}[H] 
%    \centering 
%    \includegraphics[width=0.6\textwidth]{pentabox.png} 
%\caption{The pentagon-box diagram}
%\end{figure}

%\label{sec:outlook}

\section*{Acknowledgement}
%\acknowledgments
We thank Alessandro Georgordis, Kasper Larsen and Xiaodi Li for their contributions in the early stage of designing this package. We thank Simon Badger, Johannes Henn, Roman Lee, David Kosower, Yingxuan Xu and Simone Zoia for very enlightening discussions. We express our gratitude to Nikolaos Syrrakos for extensively testing our package, as well as to Yefan Wang for providing us with valuable test examples. YZ is
supported from the NSF of China through Grant No. 11947301, 12047502, 12075234, 12247103 and
the Key Research Program of the Chinese Academy of Sciences, Grant No. XDPB15.
%should state who is supported by this grant to avoid misunderstanding
Gef\"ordert durch die Deutsche Forschungsgemeinschaft (DFG) - Projektnummer 286237555 - TRR 195 (Funded by the Deutsche Forschungsgemeinschaft (DFG, German Research Foundation) - Project-ID 286237555 - TRR 195). The work of JB was supported by Project B5 of SFB-TRR 195 and Potentialbereich \emph{SymbTools - Symbolic Tools in Mathematics and their Application} of the Forschungsinitative Rheinland-Pfalz.
ZW expresses gratitude to the other authors of this paper for their agreement on the ordering of the author names.
%\appendix

\bibliographystyle{elsarticle-num}
\bibliography{bibtex}

\end{document}